\documentclass[preprint,12pt]{article}
\usepackage{amsfonts}
\usepackage{amssymb,amsmath}
\begin{document}

\title{\begin{flushright} ${}$\\[-40pt] $\scriptstyle \mathrm SB/F/438-14$
\\[0pt]
\end{flushright}
  Electromagnetic fields in matter revisited}
\author{Rodrigo Medina \\
\small Centro de F\'{\i}sica\\[-0.8ex]
\small Instituto Venezolano de Investigaciones Cient\'{\i}ficas\\[-0.8ex]
\small Apartado 20632 Caracas 1020-A, Venezuela\\
\small \texttt{rmedina@ivic.gob.ve}\\
\and   J. Stephany\\
\small  Departamento de F\'{\i}sica\\[-0.8ex]
\small Universidad Sim\'{o}n Bol\'{\i}var\\[-0.8ex]
\small Apartado 89000, Caracas 1080A, Venezuela.\\
\small \texttt{stephany@usb.ve}
}

\date{}
\maketitle
\thispagestyle{empty}
\begin{abstract}
The force density on matter and the kinetic energy-momentum tensor of the
electromagnetic field in matter are obtained starting from  Maxwell equations
and Lorentz force  at microscopic level and averaging over a small region of
space-time. The  macroscopic force density is taken to depend linearly on the
average fields and their first derivatives and is shown to be determined by two
phenomenological fields which are subsequently identified with the  free current 
density  and the
polarization density tensor. It is shown that as expected, the average current
density is the sum of the free current density and a dipolar contribution and
that the average  field fulfill the  Maxwell equations.
The macroscopic energy-momentum tensor of field is shown to be equal to the
standard empty-space energy-momentum tensor built with the macroscopic fields
plus a dipolar correction. The density of momentum of the field is confirmed to
be given by Minkowski's expression. The energy-momentum tensor of macroscopic
matter is equal to the average of the microscopic energy-momentum tensor of
matter plus the difference between the average tensor of microscopic fields and
the macroscopic tensor of fields.
\end{abstract}

\section{Introduction}

The determination of which is the correct expression for the momentum of
photons with energy $E$ in a material media has been, for many years and
unsettle issue actively researched \cite{Brevik1979,Pfeifer2007,MilonniBoyd2010,BarnettLoudon2010}. The dispute is between the values $nE/c$
and  $E/nc$ with $n$ the refraction index, associated respectively with the
names of Minkowski and Abraham each of whom proposed at the beginning of the
last century an expression for the energy-momentum tensor
\cite{Minkowski1908,Abraham1910}. Although most of the experimental results
support Minkowski expression some theoretical arguments related with the
movement of the center of mass of matter-field systems prove to have a strong
convincing power  in the mind of many people
\cite{Balazs1953,Pfeifer2007,BarnettLoudon2010}. These
arguments are closely related to the fact that Minkowski tensor is
non-symmetric whereas Abraham's object is symmetric and are consequence of
supposing that the center of mass (energy) of a matter-field system should
always be inertial.  

The theoretical arguments supporting Abraham expression turn out to be wrong
\cite{MedandSa,MedandSb}. Recently, we observed  that  the usual forms of the
center of mass motion theorem may be violated if the energy-momentum of the
system is non-symmetric \cite{MedandSa} and that this violation occurs in some
electromagnetic systems. Motivated by this observation we  were able
\cite{MedandSb} to present the correct energy-momentum tensor for the
electromagnetic field in a material media which is consistent with the
microscopic expression  of the Lorentz force. It results to be non-symmetric.
The obtained  expression differs of Minkowski proposal in the diagonal terms
but defines the same  momentum density. In this form the controversy on the
electromagnetic momentum in material media is resolved in favor of Minkowski's
expression.   In these works, we also stressed that Abraham's object is not a
Lorentz tensor, a fact that was already pointed out in the literature
\cite{VeselagoShchavlev2010} but has not received the attention that it deserves.

In the second mentioned work, we first obtain the Lorentz invariant
expression of the force density assuming that the force on a non-polarized but
magnetized element of material is equal to the Lorentz force on a magnetic
dipole with the same magnetic moment. Then the energy-momentum tensor is
obtained by consistency with Maxwell equation and imposing that Newton's third
law between field and matter holds. In this approach we suppose that charges
can be separated into free charges and bound charges. The total bound charge of
any piece of material is always zero, so that the bound charge at the surface
is opposite to the bound charge in the bulk. This leads to the definition of a
polarization field $\mathbf{P}$ such that the densities of dipolar (bound)
charges are $\sigma_{\mathrm d}=\mathbf{P}\cdot\hat{\mathbf{n}}$ in the surface
and $\rho_{\mathrm d}=-\nabla\cdot\mathbf{P}$ in the  bulk. The current density
of bound charges must then be
 $\mathbf{j}_{\mathrm d}=\dot{\mathbf{P}}+ c\nabla\times\mathbf{M}$.

To understand better the nature of the energy-momentum tensor obtained it is
convenient to look for alternative deductions. The traditional treatment of the
phenomenological fields,  $\mathbf{P}$ and  $\mathbf{M}$ which appear in the
macroscopic Maxwell equations is to obtain those fields by making averages of
microscopic electric and magnetic dipoles. However, in this approach the
covariant treatment  presents some problems in how to define consistently the
dipole moments from microscopic charges and currents, particularly when the
dipoles are moving or changing in time \cite{JacJ1998,Groot1972}. Instead of
this in the present paper we present a third approach
which is independent of any
microscopic model of the material and does not require to use the force on a
magnetic dipole. We suppose that the macroscopic quantities may be obtained by
taking the average of microscopic fields and forces over small regions of
space-time and that the macroscopic force density may be expanded linearly in
 the macroscopic fields and their derivatives. In the dipolar approximation
only the first derivatives are relevant.  We define the free charge current
density as the four-vector that couples to the average field and the 
polarization density tensor as the parameters that couple to the first
derivatives in the expansion. Then we obtain
self-consistently the expressions for the free and bound (dipolar) charge and
current densities and for the energy-momentum tensor. Finally we show that the
polarization tensor that was defined as the coupling parameters of the force
should indeed be interpreted as the polarization-magnetization tensor.

The inconsistencies between Maxwell equations and the theoretical arguments
supporting Abraham's point of view led also to a  long lasting controversy
about which is the true force that is exerted on the dipole densities and how
the total energy-momentum tensor has to be split in the part assigned to the
macroscopic electromagnetic field and the part assigned to matter
\cite{Pfeifer2007}. Some authors even consider that such splitting may be done 
arbitrarily \cite{Pfeifer2007} or that both definitions of the photon momentum may be used meaningfully depending on the context \cite{Barnett2010}. In our opinion this discussions should be cut off once one realizes that the macroscopic force density is determined by the microscopic Lorentz force  and that the use of Newton's third law between matter and field unequivocally determines the splitting of the energy-momentum tensor.

\section{Space-time average}
At a microscopic scale, we assume that for the electromagnetic field
$F^{\mu\nu}$ and the current density $j^\mu$, Maxwell equations and Bianchi's
 identity hold
\begin{equation}
\label{Maxwell}
\partial_\nu F^{\mu\nu} = \frac{4\pi}{c} j^\mu \ ,
\end{equation}
\begin{equation}
\label{Bianchi}
\partial^\lambda F^{\mu\nu}+\partial^\mu F^{\nu\lambda}+\partial^\nu
F^{\lambda\mu}=0 \ ,
\end{equation}
and that the force density is given by Lorentz expression 
\begin{equation}
\label{Lorentz}
f^\mu =\frac{1}{c}F^\mu_{\ \alpha}j^\alpha\ .
\end{equation}
We use the metric
$\eta^{\mu\nu}= \hbox{diag}(-1,1,1,1)$ and Gauss units. Correspondingly the
energy-momentum tensor for the microscopic electromagnetic field is the
standard tensor given by 
\cite{LanLL1994},
\begin{equation}
\label{TEI-S}
T^{\mu\nu}_{\mathrm S}=-\frac{1}{16\pi}\eta^{\mu\nu}F^{\alpha\beta}F_{\alpha\beta}
+\frac{1}{4\pi}F^\mu_{\ \alpha}F^{\nu\alpha}\ .
\end{equation}
which satisfy identically $\partial_\nu T^{\mu\nu}_{\mathrm S}=-f^\mu$ as a
 consequence of
(\ref{Maxwell}), (\ref{Bianchi}) and (\ref{Lorentz}). The
force density can alternatively be obtained from the energy-momentum tensor of
 matter by
$f^\mu = \partial_\nu T^{\mu\nu}_{\mathrm M}$ if it is available.

The observable macroscopic quantities correspond to averages of the
microscopic quantities over small regions of space-time in the form,
\begin{equation}
\label{average}
\langle A(x)\rangle = \int A(x^\prime)W(x-x^\prime)\,dx^\prime\ ,
\end{equation}
where $W(x)$ is a smooth function that is essentially constant inside a region
of size $R$ and that vanishes outside
\begin{equation}
W(x)\ge 0\ ,\qquad |x^\mu|>R \Longrightarrow W(x)=0\ ,\qquad \int W(x)\,dx =1\ .
\end{equation}
Inside the region $W(x)\approx \langle W\rangle$.

At the scale $R$ the microscopic fluctuations are washed out,
and  therefore all the products of averages and fluctuations are
negligible. That is, if $\delta A= A-\langle A\rangle$ then
$\langle \langle B\rangle\delta A\rangle\approx 0$ and also
 $\langle\langle B\rangle\rangle \approx \langle B\rangle$.

With  these conditions it follows that
\begin{eqnarray}
\label{average_derivative}
 \partial_\nu \langle A\rangle &=& \int A(x^\prime)\partial_\nu W(x-x^\prime)
\, dx^\prime=
-\int A(x^\prime)\partial^\prime_\nu W(x-x^\prime)\,dx^\prime\nonumber\\ 
&=&-\int\partial^\prime_\nu[A(x^\prime)W(x-x^\prime)]\,dx^\prime+
\int \partial^\prime_\nu A(x^\prime)W(x-x^\prime)\,dx^\prime\nonumber\\
&=& \langle \partial_\nu A\rangle \ .
\end{eqnarray}

We denote the macroscopic fields with a bar. In particular,
$\bar{F}^{\mu\nu} =\langle F^{\mu\nu}\rangle$. Using (\ref{average_derivative})
one immediately obtains that Maxwell equations (\ref{Maxwell}) are valid for
macroscopic fields and the averaged currents
\begin{equation}
\label{Maxwell_average}
\partial_\nu\bar{F}^{\mu\nu}=\frac{4\pi}{c}\langle j^\mu\rangle
\end{equation}
and that Bianchi's identity is valid for $\bar{F}^{\mu\nu}$ and also for
$\delta F^{\mu\nu}=F^{\mu\nu}-\bar{F}^{\mu\nu}$.

\section{Macroscopic force}

As mentioned above, reported apparent inconsistencies in the description of the
electromagnetic field in material media led some authors to propose alternative
forms of the force which the electromagnetic field exerts on a material media.
These where pioneered by Einstein and Laub \cite{EinsteinLaub} in 1907 although
it seems that some years later Einstein retracted of the suggestion
\cite{Walter}. These proposals have problems either with Lorentz invariance or
with being quadratic in the magnetization. The related hypothesis of the
existence of a hidden momentum \cite{SJ1967} has also problems with the
mechanical interpretation of its behavior. In this letter we avoid any
controversy about this issue by making the sole hypothesis that the microscopic
force is given by the Lorentz expression. From this we obtain a definite
expression of the force density in the macroscopic description which depends on
a phenomenological tensor $D^{\mu\nu}$ whose components will be later on
identified with the magnetization and polarization vectors.  Since we are
supposing that momentum is conserved,  the matter and field portions of the
energy-momentum tensor may be the characterized by supposing that Newton's third
law holds between field and matter.

In order to obtain the average current density and the macroscopic
energy-momentum tensors of fields and matter we compute first the average of
the microscopic forces. Since the force (\ref{Lorentz}) is quadratic in the
microscopic quantities the result in this case is not straightforward. We
assume that the macroscopic force density $\bar{f}^\mu$ on matter shares with
the microscopic force density 
the following four properties that survive the averaging process. 1) It is a
local functional of the average field.
 2) It is linear in the field. 
 3) The direction of the force should be determined by the field. 4) It
transforms as a four vector.
Therefore the force density can be expanded as a sum of terms in
which the derivatives of the field tensor are contracted with matter-dependent
tensors and the free index belongs to the field derivative tensors.
 In the dipolar approximation,
 we neglect the coupling with second and higher order
derivatives which corresponds to quadrupolar and higher order couplings. Taking into account Bianchi's identity, it can be proved that the most general
expression for the macroscopic force density that fulfills the four stated
above conditions takes the form
\begin{equation} 
\label{Mac-force}
\bar{f}^\mu = \frac{1}{c}\bar{F}^\mu_{\ \alpha}j^\alpha_{\mathrm f}+
\frac{1}{2}\partial^\mu \bar{F}^{\alpha\beta} D_{\alpha\beta}\ .
\end{equation}
It is written in terms of two phenomenological quantities related to matter,
a four-vector $j^\alpha_{\mathrm f}$ and an antisymmetric four-tensor
$D_{\alpha\beta}$ to be identified later in a consistent way respectively with the current density of
free charges and the dipolar tensor density. Note that then the temporal component of 
$D_{\alpha\beta}$ will be identified with the polarization vector $\mathbf{P}$, $P_k = D_{0k}$, 
and the spatial components will be in correspondence 
with the magnetization vector $\mathbf{M}$  by $D_{ij}=\epsilon_{ijk}M_k$.

With the use of some simple algebra and Bianchi's identity the force expression
takes the alternative form
\begin{equation}
\label{Mac-force-2}
\bar{f}^\mu = \frac{1}{c}\bar{F}^\mu_{\ \alpha}(j^\alpha_{\mathrm f}+
c\partial_\beta D^{\alpha\beta})+
\partial_\beta(\bar{F}^\mu_{\ \alpha} D^{\beta\alpha})\ .
\end{equation}
In the microscopic formulation the total energy momentum tensor $T^{\mu\nu}$
splits naturally in a contribution of the field given by (\ref{TEI-S}) and a
contribution $T^{\mu\nu}_{\mathrm M}$ of the matter. The average of the total
microscopic energy-momentum tensor is the macroscopic energy momentum tensor,
but how it splits in a macroscopic matter term $\bar{T}^{\mu\nu}_{\mathrm M}$
and a macroscopic field term $\bar{T}^{\mu\nu}_{\mathrm F}$ should be
determined by Newton's third law. So we write,
\begin{equation}
\bar{T}^{\mu\nu}_{\mathrm F} + \bar{T}^{\mu\nu}_{\mathrm M} =
 \langle  T^{\mu\nu}_{\mathrm S} + T^{\mu\nu}_{\mathrm M}\rangle \ .
\end{equation}
and impose,
\begin{equation}
\label{NewtonLaw}
\partial_\beta \bar{T}^{\mu\beta}_{\mathrm M}=\bar{f}^\mu = -\partial_\beta \bar{T}^{\mu\beta}_{\mathrm F} .
\end{equation} 
In particular the average of the microscopic energy-momentum tensor of
electromagnetic fields is not the macroscopic energy-momentum of the field. It
may be expressed as, 
\begin{equation}
\langle T^{\mu\nu}_{\mathrm S}\rangle = \bar{T}^{\mu\nu}_{\mathrm S}+
\langle \delta T^{\mu\nu}_{\mathrm S}\rangle\ , 
\end{equation}
where $\bar{T}^{\mu\nu}_{\mathrm S}$ is the standard tensor built with the
macroscopic fields and $\delta T^{\mu\nu}_{\mathrm S}$ is the standard tensor
built with the fluctuation fields $\delta F^{\mu\nu}$. For the actual
macroscopic energy-momentum tensor of fields inside matter we write,
\begin{equation}
\bar{T}^{\mu\nu}_{\mathrm F} = \bar{T}^{\mu\nu}_{\mathrm S} + \Delta T^{\mu\nu}_{\mathrm F} \ ,
\end{equation} 
where $\Delta T^{\mu\nu}_{\mathrm F}$ is a possible polarization-dependent
correction that will be determined self-consistently. The terms linear in the
fluctuation were neglected as explained above. To obtain the macroscopic
energy-momentum tensor of matter, the average fluctuation tensor of the field
should be included and the dipolar correction must be subtracted,
\begin{equation}
\bar{T}^{\mu\nu}_{\mathrm M} = \langle T^{\mu\nu}_{\mathrm M}\rangle + 
\langle \delta T^{\mu\nu}_{\mathrm S}\rangle - \Delta T^{\mu\nu}_{\mathrm F}\ . 
\end{equation}
Note that the fluctuation fields $\delta F$ should be considered part of the macroscopic matter.
This is the reason why $\bar{f}^\mu \not=\langle f^\mu\rangle$.

Using (\ref{NewtonLaw}), the macroscopic force density is
\begin{equation}
\label{Mac-force-3}
\bar{f}^\mu =\partial_\beta \bar{T}^{\mu\beta}_{\mathrm M}=\langle f^\mu\rangle+
\partial_\beta[\langle \delta T^{\mu\beta}_{\mathrm S}\rangle - \Delta T^{\mu\beta}_{\mathrm F}]\ .
\end{equation} 
Using (\ref{Maxwell}) and (\ref{Lorentz}) the average microscopic
force density can be expressed as
\begin{equation}
\langle f^\mu\rangle =\frac{1}{4\pi}\langle F^\mu_{\ \alpha}\partial_\beta
F^{\alpha\beta}\rangle\ .
\end{equation}
Making the substitution $F^{\alpha\beta}=\bar{F}^{\alpha\beta}+
\delta F^{\alpha\beta}$ one gets
\begin{equation}
\langle f^\mu\rangle =\frac{1}{4\pi}\bar{F}^\mu_{\ \alpha}\partial_\beta
\bar{F}^{\alpha\beta}+ \frac{1}{4\pi}\langle\delta F^\mu_{\ \alpha}
\partial_\beta\delta F^{\alpha\beta}\rangle\ .
\end{equation}
Using (\ref{Maxwell_average}) in the first term of this last equation and
Bianchi's identity in the second, the average microscopic force density is then
\begin{equation}
\langle f^\mu\rangle =\frac{1}{c}\bar{F}^\mu_{\ \alpha}\langle j^\alpha\rangle
-\langle\partial_\beta\delta T^{\mu\beta}_{\mathrm S}\rangle\ .
\end{equation}
Substituting this result in (\ref{Mac-force-3}) and equating with
 (\ref{Mac-force-2}) one finally gets
\begin{equation}
\label{condition}
\frac{1}{c}\bar{F}^\mu_{\ \alpha}(j^\alpha_{\mathrm f}+c\partial_\beta
 D^{\alpha\beta}-\langle j^\alpha\rangle)=-\partial_\beta(\bar{F}^\mu_{\ \alpha}
D^{\beta\alpha}+\Delta T^{\mu\beta}_{\mathrm F})\ .
\end{equation}
This last equation is satisfied identically by setting
\begin{equation}
\label{current_density}
\langle j^\alpha\rangle = j^\alpha_{\mathrm f}+ c\partial_\beta D^{\alpha\beta}
\end{equation}
and 
\begin{equation}
\label{correction}
\Delta T^{\mu\nu}_{\mathrm F} = -\bar{F}^{\mu}_{\ \alpha}D^{\nu\alpha}\ .
\end{equation}
These solutions  are unique. In fact, since (\ref{condition}) must be satisfied
for any field tensor, equation  (\ref{current_density}) follows. A tensor
$X^{\mu\nu}$ whose divergence
vanish identically ($\partial_\nu X^{\mu\nu}\equiv 0$) could be added to 
$\Delta T^{\mu\nu}_{\mathrm F}$, but since $\bar{T}^{\mu\nu}_{\mathrm F}$ can
be at most quadratic in the fields, and should vanish when the fields vanish
such a tensor also vanishes.

Defining the dipolar current density
\begin{equation}
 j^\alpha_{\mathrm d}=c\partial_\beta D^{\alpha\beta}.
\end{equation}
Maxwell equations for the macroscopic fields are written in the familiar form,
\begin{equation}
\label{Maxwell_dipolar}
\partial_\nu\bar{F}^{\mu\nu}=
 \frac{4\pi}{c}(j^\mu_{\mathrm f}+j^\mu_{\mathrm d})\ .
\end{equation}
Noting that in  terms of $\mathbf{P}$ and $\mathbf{M}$ defined above, the also familiar expressions 
$\rho_{\mathrm d}= -\nabla\cdot\mathbf{P}$  and $\mathbf{j}_{\mathrm d}= \dot{\mathbf{P}}+
c\nabla\times\mathbf{M}$ are obtained, the identification of $j^\alpha_{\mathrm f}$ as the current of free charges and $D^{\nu\alpha}$ with the dipolar density is almost completed.

The dipolar or bound charge is conserved identically, $\partial_\alpha j^\alpha_{\mathrm d}=c\partial_\alpha\partial_\beta D^{\alpha\beta}=0$. Since the average charge is conserved,
equation (\ref{current_density}) implies that the free charge is also conserved,
$\partial_\alpha j^\alpha_{\mathrm f}=0$.

It is worth noting that this whole scheme leaves out
processes, like ionization or capture of carriers, in which there is exchange
between free charges and bound charges.  In those cases (\ref{Mac-force}) does
not hold. 

Now we proceed to discuss the energy-momentum tensor. It is convenient to define the tensor
\begin{equation}
H^{\alpha\beta} = \bar{F}^{\alpha\beta} - 4\pi D^{\alpha\beta}\ ,
\end{equation}
which is related to the electric displacement $\mathbf{D}=\mathbf{E}+4\pi\mathbf{P}$
and the magnetizing field $\mathbf{H}=\mathbf{B}-4\pi\mathbf{M}$, $H^{0k}=D_k$ and
$H^{ij}=\epsilon_{ijk}H_k$. With this tensor Maxwell equations (\ref{Maxwell}) become
\begin{equation}
\partial_\nu H^{\mu\nu}=\frac{4\pi}{c}j^\mu_{\mathrm f}
\end{equation}
and the macroscopic energy-momentum tensor of fields is
\begin{equation}
\bar{T}_{\mathrm F}^{\mu\nu}=\bar{T}_{\mathrm S}^{\mu\nu} + \Delta T^{\mu\nu}_{\mathrm F}=
-\frac{1}{16\pi}\eta^{\mu\nu}\bar{F}^{\alpha\beta}\bar{F}_{\alpha\beta}+
\frac{1}{4\pi}\bar{F}^\mu_{\ \alpha}H^{\nu\alpha}\ . 
\end{equation}
This is exactly the result we obtained in Ref.\cite{MedandSb}

\section{Densities of dipole moment}
To  completely recover the usual picture let us show that 
$\mathbf{P}$ and $\mathbf{M}$ are the densities of electric dipole moment and of
magnetic dipole moment respectively without relying in the model of microscopic dipoles. 

First we  find  the charges  at the surface of a piece of material. Near the surface
$\mathbf{P}$, go smoothly to zero in a distance of the order of $R$. The total dipolar charge is
obtained by integrating $\rho_{\mathrm d}=-\nabla\cdot\mathbf{P}$ over the volume of the material. Because of Gauss' theorem such integral is zero since $\mathbf{P}=0$ at the surface. Then the total dipolar charge is always zero. It is a bound charge  that cannot leave the material. At a macroscopic scale (much bigger than $R$) there is a discontinuity at the surface. The dipolar charge at the surface is always opposite to the charge in the bulk.  If $\sigma_{\mathrm d}$ is the surface
density of polarization, then
\begin{equation}
0=\oint \sigma_{\mathrm d}\,dS-\int \nabla\cdot\mathbf{P}\,dV=
\oint[\sigma_{\mathrm d}-\mathbf{P}\cdot\hat{\mathbf{n}}\,]\,dS\ .
\end{equation}
This can happen for any surface if
\begin{equation}
\sigma_{\mathrm d} =\mathbf{P}\cdot\hat{\mathbf{n}}\ ,
\end{equation} 
which of course is the usual expression given by the model of microscopic dipoles.
The electrical dipole moment of a piece of material is computed as always giving
\begin{eqnarray}
\mathbf{d}&=&\oint \mathbf{x}\sigma_{\mathrm d}\,dS +\int\mathbf{x}\rho_{\mathrm d}\,dV=\int\mathbf{P}\,dV\ 
\end{eqnarray}
This completes the interpretation of $\mathbf{P}$ as the density of dipole moment.

The magnetic current density in the bulk is
$\mathbf{j}_{\mathrm M}=c\nabla\times\mathbf{M}$. In addition there is also a surface
current density $\mathbf{\Sigma}_{\mathrm M}$ that
can be obtained in a way similar to that used for $\sigma_{\mathrm d}$. Let us
consider a closed
curve that is outside the piece of material and is the border of a surface
that cuts the material.
The total magnetic current that crosses
the surface is the flux of $\mathbf{j}_{\mathrm M}$. Because of Stokes' theorem
such flux is the circulation of $c\mathbf{M}$ in the curve which is zero.
Therefore for any surface $S$ that cuts the material the total magnetic current
that crosses the surface is zero. At a macroscopic scale the bulk current is
opposite to the surface current. Let ${\cal C}$ be the
intersection of $S$ with the surface of the piece of material, 
 $\hat{\mathbf{t}}$ the unitary vector tangent to the curve ${\cal C}$ and
$\hat{\mathbf{n}}$ the unitary vector orthogonal to the surface of the piece. The
unitary vector which is orthogonal to ${\cal C}$ and tangent to the surface of the
piece is $\hat{\mathbf{n}}\times\hat{\mathbf{t}}$. Then
\begin{eqnarray}
0&=&\oint_{\cal C}\mathbf{\Sigma}_{\mathrm M}\cdot\hat{\mathbf{n}}\times\hat{\mathbf{t}}\,dl
+\int_S c\nabla\times\mathbf{M}\cdot d\mathbf{S}\nonumber\\
&=&\oint_{\cal C}[\mathbf{\Sigma}_{\mathrm M}\cdot\hat{\mathbf{n}}\times\hat{\mathbf{t}}+
c\mathbf{M}\cdot\hat{\mathbf{t}}\,]dl\nonumber\\
&=&\oint_{\cal C}[\mathbf{\Sigma}_{\mathrm M}\times\hat{\mathbf{n}}+c\mathbf{M}]\cdot
\hat{\mathbf{t}}\,dl\ .
\end{eqnarray}
The expression that fulfills this equation for any $\hat{\mathbf{t}}$ is
\begin{equation}
\mathbf{\Sigma}_{\mathrm M}= c\mathbf{M}\times\hat{\mathbf{n}}\ .
\end{equation}
The usual computation of the magnetic dipole moment of the magnetic currents, 
\begin{eqnarray}
\mathbf{m}&=&\frac{1}{2c}\oint \mathbf{x}\times\mathbf{\Sigma}_{\mathrm M}\,dS
+\frac{1}{2c}\int\mathbf{x}\times\mathbf{j}_{\mathrm M}\,dV=
\int\mathbf{M}\,dV\ .
\end{eqnarray}
completes the interpretation of $\mathbf{M}$ as the density of
magnetic dipole moment.

\section{Conclusion}

Recently we have presented a deduction of the  energy momentum-tensor of the
electromagnetic field in matter \cite{MedandSb} that should put an end to the
long dated Abraham-Minkowski controversy. Our result confirms Minkowski's
expressions
$\mathbf{g}_{\mathrm{Min}} = \mathbf{D}\times\mathbf{B}/4\pi c$ for
 the momentum density of the field in matter and  
$p=nE/c$ for the momentum of a photon of energy $E$, $n$ being the refraction
index. Moreover our result also predicts that an electromagnetic wave incident
on a dielectric block will pull the block instead of pushing it. This
invalidates Balazs \cite{Balazs1953} famous argument in favor of Abraham
momentum and has the simple physical explanations that dielectric tends to move
in the direction of higher field. Due to the many arguments that have been
discussed along the years on this issue and also to get a better understanding
of the nature of result obtained we found convenient to look for an alternative
deduction of the energy-momentum tensor that we report in this paper. 

By taking averages over small regions of space-time we have obtained the laws
of macroscopic electromagnetism, including the expressions for the force and
the energy-momentum tensor that previously were
in dispute  for a long time \cite{Pfeifer2007}. Our derivation is independent of
any particular model of microscopic matter. We show that supposing the validity
of the Lorentz force at the microscopic level, there is a unique result which
is compatible with the validity of Maxwell equations and Bianchi's identity,
both at the microscopic and the macroscopic levels, and for which the
macroscopic force is linear in the macroscopic field and its derivatives.

In our approach the polarization tensor is introduced as a phenomenological
tensor which couples with the field derivatives in the expansion of the force
expression. Then the dipolar current density and the energy-momentum tensor are
determined by consistency. Finally it is shown that the polarization tensor can
be  indeed interpreted as the density of dipolar moments.

The force on the dipoles is not the same as the force
on the dipolar charges and currents. To see why this is true, consider a
piece of material subdivided into infinitesimal elements. For each element the
dipolar moments are determined by the surface charges and currents, and so it is
the force. The total force on the piece is the sum of the forces on each
element, that is, it can be calculated by integrating the force density (9).
In contrast the contributions to the field due to the surface charges and
currents of adjacent elements cancel out; what remains are the contributions of
the bulk and of the external surface of the piece of material. The present
treatment shows that it is inconsistent to assume, as it has been proposed for example in
\cite{Obukhov2003},  that the force on a polarized material is the Lorentz
force on the total current. If the force is of the Lorentz kind then it must be that
$\mathbf{P}=\mathbf{M}\equiv0$.

Our results for the force density and the energy-momentum tensor agree exactly
with those  reported by us in Ref.~\cite{MedandSb}, with the
methodological advantage that in the present derivation we do not make use of
the expression for the force density  on the matter dipoles.

Our expression for the energy density  
\begin{equation}
\label{energyd}
u=\bar{T}^{00}_\mathrm{F}=\frac{1}{8\pi}(E^2+B^2)+\mathbf{P}\cdot\mathbf{E}
\end{equation} 
is different from the energy density of  fields without dipolar coupling. The
difference $\mathbf{P}\cdot\mathbf{E}$ is the opposite of the electrostatic
energy density of the polarization. This energy is therefore subtracted from
the energy of the field and considered part of the energy of matter. Note that
there is no similar magnetic term, since no potential magnetic energy exist.
Equation (\ref{energyd})  is also different from Poynting's expression, 
 $u_\mathrm{P}=(\mathbf{E}\cdot\mathbf{D}+\mathbf{B}\cdot\mathbf{H})/8\pi$, but
our energy current density is the Poynting vector
 $\mathbf{S}=c\mathbf{E}\times\mathbf{H}/4\pi$. The energy
conservation equation in our formulation
\begin{equation}
\frac{\partial u}{\partial t}+\nabla\cdot\mathbf{S}=-\mathbf{E}\cdot
\mathbf{j}_\mathrm{f}+\mathbf{P}\cdot\frac{\partial\mathbf{E}}{\partial t } +
\mathbf{M}\cdot\frac{\partial\mathbf{B}}{\partial t }
\end{equation}
differs also from  Poynting conservation equation  which is given by
\begin{equation}
 \frac{\partial u_\mathrm{P}}{\partial t} +\nabla\cdot\mathbf{S} =-\mathbf{E}\cdot\mathbf{j}_\mathrm{f}\ .
\end{equation}

Our results are valid for any kind of material in any kind of condition
(except when there is exchange between free and bound charges): ferromagnets,
saturated paramagnets, electrets, matter moving or at rest, solids, fluids,
absorbing and dispersive media, etc.
  Poynting's equation is obtained from our result in the particular
case of linear polarizabilities. In such a case the dipolar contributions to
the power density on the electromagnetic field
can be integrated, yielding the opposite of the energy density increase of
matter when it is polarized. Such an energy increase should be added to the
pure field energy density to obtain Poynting's energy density. Therefore,
Poynting's energy density corresponds to a mixed entity composed of ``fields
plus polarizations''. On the other hand Minkowski's is the only relativistic
tensor for which
$T^{00}=u_\mathrm{P}$ and $cT^{0k}=S^k$ in any frame of reference. For an
electromagnetic wave propagating in a medium it makes sense to include the
polarization energy of matter as part of the wave energy. So Minkowski's
tensor properly represents the energy-momentum tensor of the EM wave in a
non-dispersive medium. It can be useful for calculating the theoretical force on the wave, but it cannot be
used for determining the force on matter, since the wave energy has a
component which belongs to the matter. For doing that one has to use our
tensor. Nevertheless, note that in general, only the force on matter can be
measured. Magnetostriction and electrostriction are  also good examples of the advantages of our approach. They are not described by Minkowski's tensor \cite{Brevik1979}, but may be calculated with our tensor. 

Poynting's equation does not hold when the polarizabilities are not linear. For
example it is well known that this is what occurs in the case of optical
dispersive media \cite{JacJ1998,Philbin2011}. Our result should be a good
starting point for a fresh approach to study such cases.  

%%%%%% References %%%%%%%%%%


\begin{thebibliography}{long}
\bibitem{Brevik1979}I.~Brevik, Phys.Rep.\textbf{52}, 133--201 (1979).
\bibitem{Pfeifer2007}Robert N.~C.~Pfeifer {\it et al}, Rev.~Mod.~Phys.
\textbf{79}, 1197--1216 (2007).
\bibitem{MilonniBoyd2010} Peter W.~Milonni and Robert W.~Boyd, Advances in
Optics and Photonics \textbf{2}, 519--553 (2010).
\bibitem{BarnettLoudon2010} Stephen M.~Barnett and Rodney Loudon,
 Phil.~Trans.~R.~Soc. A  \textbf{368}, 927--939 (2010).
\bibitem{Minkowski1908} H.~Minkowski, Nachr.~Ges.~Wiss.~G\"ottingen, 53 (1908).
\bibitem{Abraham1910} M.~Abraham, Rend.~Circ.~Mat.~Palermo \textbf{30}, 33
(1910).
\bibitem{Balazs1953} N.~L.~Balazs, Phys.~Rev. \textbf{91}, 408--411 (1953).
\bibitem{MedandSa} Rodrigo Medina and J.~Stephany, {\it Violation of the center of mass theorem for systems with electromagnetic interaction},  arXiv:1404.5251. 
\bibitem{MedandSb} Rodrigo Medina and J.~Stephany, Rodrigo Medina and J.~Stephany, {\it The force density and the kinetic energy-momentum tensor of electromagnetic fields in matter},  arXiv:1404.5250.
\bibitem{VeselagoShchavlev2010} V.~G.~Veselago, V.~V.~Shchavlev,
Physics--Uspekhi \textbf{53}, 317--318 (2010).
\bibitem{JacJ1998} J.~D.~Jackson, {\it Classical Electrodynamics 3rd ed.},
John Wiley\&Sons, New York, 1998, p. 248--258.
\bibitem{Groot1972}S.~R.~de Groot and L.~G.~Suttorp, {\it Foundations of 
Electrodynamics}, North-Holland, Amsterdam, 1972, Ch V. 
\bibitem{Barnett2010}Stephen M.~Barnett, Phys.~Rev.~Lett. \textbf{104}, 070401 (2010).
\bibitem{LanLL1994} L.~Landau and E.~Lifchitz, {\it Classical theory of fields},
Butterworth-Heinemann, Oxford, 1994, p. 87.
\bibitem{EinsteinLaub} A. Einstein and J. Laub,  Ann. Phys.  \textbf{26},
 541--550 (1908).
\bibitem{Walter}S.~Walter, Minkowski, mathematicians, and the mathematical
 theory of relativity, in {\it The Expanding Worlds of General Relativity},
 H.~Goenner, ed.  (Birkhäuser, 1999), pp. 45--86.
\bibitem{SJ1967} W.~Shockley and R.P.~James, Phys.~Rev.~Lett. \textbf{18},
 876--879 (1967).
\bibitem{Obukhov2003}Yu.~N.~Obukhov and F.~W.~Hehl, Physics Letters A
 \textbf{311}, 277--284, (2003).
\bibitem{Philbin2011} T.~G.~Philbin,  Phys. Rev. A \textbf{83}, 013823 (2011).
\end{thebibliography}
\end{document}